\documentclass[12pt]{article}

\usepackage[toc,page]{appendix}
\usepackage{slashed}
\usepackage[sumlimits,intlimits,namelimits]{amsmath} 
\usepackage[english]{babel}
\usepackage{graphicx}
\newcommand{\be}{\begin{equation}}
\newcommand{\ee}{\end{equation}}
\newcommand{\ba}{\begin{eqnarray}}
\newcommand{\ea}{\end{eqnarray}}
\newcommand{\nn}{\nonumber}

\newcommand{\ice}[1]{\relax}

\newcommand{\MSbar}{\overline{\rm MS}}

\usepackage{color}
\usepackage{amsmath}
\usepackage{booktabs} 
\usepackage{subfigure}

\begin{document}

\begin{titlepage}

\begin{flushright}
SI-HEP-2019-10\\
SFB-257-P3H-19-022
\end{flushright}
\begin{center}
  { \Large\bf 
    QCD corrections to 
    inclusive heavy hadron weak decays\\ at
    \boldmath $\Lambda_{\rm QCD}^3 /m_Q^3$ \unboldmath}
\end{center}
\begin{center}
{\sc Thomas Mannel } and {\sc Alexei A. Pivovarov} \\[0.1cm]
{\sf Theoretische Elementarteilchenphysik, Naturwiss.- techn. Fakult\"at, \\
Universit\"at Siegen, 57068 Siegen, Germany}
\end{center}


\begin{abstract}\noindent
  We present an analytical calculation of the $\alpha_s$ corrections 
  for the coefficient of $\rho_D/m_Q^3$ term in the heavy quark expansion   
  for the inclusive semileptonic decays of heavy hadrons, such as 
  $B \to X_c \ell \bar{\nu}$. The full dependence of the coefficient 
  on the final-state quark mass is taken into account.
  Our result leads to further improvement of 
  the theoretical predictions for the precision determination of
  CKM matrix element $|V_{cb}|$.  
\end{abstract}

\end{titlepage}

\section{Introduction
  \label{sec:Intro}}
The discovery of a Higgs boson a few years ago
completed the the Standard Model of
particle physics (SM), which thereby became a highly predictive framework, i.e. it 
allows us to perform very precise calculations. On the experimental side there is 
currently no hint at any particle or interaction which is not described by the SM,  
even at the highest possible energies. This implies that particle physics is about 
to enter an era of precision measurements of the SM parameters. 

In particular, accurate measurements accompanied by precise theoretical calculations
in the flavour sector of the SM have already proven to have an enormous reach at 
scales that are much larger than the center-of-mass of any existing or projected 
colliders~\cite{Charles:2015gya}. Aside from large-scale experimental efforts, this 
strategy also requires accurate theoretical computations.

The need in obtaining a high precision of theoretical predictions in particular in 
the flavour sector is urgent, since the structure of the quark  mixing is expected 
to be rather sensitive to possible the effects from physics beyond the SM (BSM). 
While the SM has successfully passed a variety of tests within current precision
(as a review, see e.g.~\cite{Butler:2013kdw,Bevan:2014iga,Forte:2015cia}),
any further insights will require the use of even more accurate theoretical
predictions.

The weak decays of quarks mediated by charged currents occur at a tree level 
and are believed to not have sizable contributions from BSM Physics. However, the 
study of such decays is importance for the precise determination of the numerical 
values of the SM parameters, in particular CKM matrix elements. For heavy quarks 
(i.e. for heavy hadrons) a reliable theoretical treatment of weak decays is possible,  
because the mass $m_Q$ of the decaying heavy quark constitutes a perturbative
scale that is much larger than the QCD infrared scale
$\Lambda_{\rm QCD}$, $m_Q\gg \Lambda_{\rm QCD}$.

Heavy quark expansion (HQE) techniques provide a systematic  
expansion of physical observables in powers of the small parameter
$\Lambda_{\rm QCD}/m_Q$. Quantitatively, the techniques work well for bottom
quarks, since a typical hadronic scale associated with binding
effects in QCD is $\Lambda_{\rm QCD} \sim 400-800~{\rm MeV}$.
With some reservations, the HQE has been used for charmed quarks as
well though the analysis is expected to be more of qualitative nature, since 
the charm-quark mass is not sufficiently large. 
Thus, the HQE and the corresponding effective theory of heavy quarks and
soft gluons (HQET) have become the major tools of modern precision analyses 
in heavy quark flavor 
physics~\cite{Shifman:1984wx,Georgi:1990um,Neubert:1993mb,Manohar:2000dt}. 

In particular for the determination of $V_{cb}$ from inclusive $b \to c$ semileptonic 
transitions the HQE has brought an enormous progress. The HQE
expansion for the total 
rate and for spectral moments have been driven to such a high accuracy that the 
theoretical uncertainty in the determination of $V_{cb}$ is now believed to be at the 
order of about one percent. However, this assumes that higher order terms in 
$\Lambda_{\rm QCD}/m_Q$ and $\alpha_s (m_Q)$ are of the expected size. In fact, 
the leading order terms (i.e. the partonic rate) has been fully computed to order  
$\alpha_s^2 (m_Q)$, the first subleading terms of order  $(\Lambda_{\rm QCD}/m_Q)^2$ 
are known to ${\cal O} (\alpha_s (m_Q))$ while all higher-order term in the 
$(\Lambda_{\rm QCD}/m_Q)^n$ ($n = 3,4,5$) are known only at tree level.    

In the present paper we analytically compute parts of the QCD corrections to the 
contributions of order $(\Lambda_{\rm QCD}/m_Q)^3$, which is one of the not yet known pieces. 
We point out that the size of these terms is expected to be of the same order as 
the partonic $\alpha_s^3 (m_Q)$ contributions, likewise the terms of order 
$\alpha_s^2 (m_Q) (\Lambda_{\rm QCD}/m_Q)^2 $. However, at the level of the current 
precision these terms turn out to be small and hence their calculation is to validate the 
assumption that they are of the expected size.  

Specifically we compute the coefficient of the power suppressed dimension six
Darwin term $\rho_D$ at next-to-leading order (NLO) of the strong coupling  
perturbation theory with the full dependence on the final state quark mass.
Compared to the calculations
at lower orders in the HQE, it has some new features
since the mixing of operators of different dimensionality in HQET has
to be taken into 
account for the proper renormalization of the Darwin term coefficient.

\section{\label{sec:rate}
Heavy Quark Expansion for Heavy Flavour Decays} 
In this section we set the stage by 
giving the very basics for the theoretical descriptions of semileptonic
decays, in particular for the decay $B\to X_cl\nu$. A more detailed description 
can be found in e.g.~\cite{Bigi:1993fe}.

The low-energy effective Lagrangian ${\cal{L}}_{\rm eff}$
for the semileptonic $b \to c l\bar{\nu}_l$ transitions reads 
\begin{equation}\label{eq:Fermi_lagr}
{\cal L}_{\rm eff} = 2\sqrt{2}{\rm G_F}V_{cb}(\bar{b}_L \gamma_\mu c_L) 
(\bar{\nu}_L \gamma^\mu \ell_L) + {\rm h.c.} \, , 
\end{equation} 
where the subscript $L$ denotes the left-handed projection of the  fermion fields and $V_{cb}$ is the relevant CKM matrix element. 

Using optical theorem one obtains the inclusive decay rate
$B\to X_c\ell\bar{\nu}_\ell$ from taking an absorptive part of 
the forward matrix element of the leading order transition operator
${\cal T}$~(see e.g.~\cite{Bigi:1993fe})
\begin{equation}\label{eq:trans_operator}
{\cal T} = i\!\!\int \! dx\,    
T\left\{ {\cal L}_{\rm eff} (x)  {\cal L}_{\rm eff} (0) \right\} \, ,
\quad \Gamma (B \to X_c \ell \bar{\nu}_\ell)
\sim \text{Im} \langle B|{\cal T} |B\rangle .
\end{equation} 
The transition operator ${\cal T}$
is a non-local functional of the quantum fields participating in the
decay process. Since the quark mass is a large scale compared to the 
scale $\Lambda_{\rm QCD}$ of QCD $m_Q\gg\Lambda_{\rm QCD}$, the relevant 
forward matrix element still contains perturbatively calculable 
contributions. These can be separated form the non-perturbative
pieces by employing effective field theory tools, which allows us 
an efficient separation of the kinematical $m_Q$ and dynamical 
$\Lambda_{\rm QCD}$ scales involved in the decay process. 

For a heavy hadron with the momentum $p_H$ and the mass $M_H$, a large part
of the heavy-quark momentum $p_Q$ is due to a pure kinematical contribution due to 
its large mass $p_Q=m_Q v+\Delta$ with $v=p_H/M_H$ being the velocity of
the heavy hadron. The momentum $\Delta$ describes the soft-scale fluctuations 
of the heavy quark field near its mass shell originating from the 
interaction with light quarks and gluons in the hadron. This is implemented 
by re-defining the heavy quark field $Q(x)$ by separating   
a ``hard'' oscillating phase and a ``soft'' field  $b_v(x)$ 
with a typical momentum of order $\Delta\sim \Lambda_{\rm QCD}$
\begin{equation}\label{eq:heavy-quark-me-phase}
Q(x) = e^{-i m_Q(v x)}b_v(x)\, .
\end{equation}
Inserting this into (\ref{eq:trans_operator}) we get 
\begin{equation}\label{eq:trans_operator1}
{\cal T} = i\!\!\int \! dx\,    e^{i m_Q v \cdot x} 
T\left\{ \widetilde{\cal L}_{\rm eff} (x)  \widetilde{\cal L}_{\rm eff} (0) \right\} \, ,
\end{equation}
where $\widetilde{\cal L}$ is the same expression as ${\cal L}$ with the replacement 
$Q(x) \to b_v(x) $. 
This makes the dependence of the decay rate on the
heavy quark mass $m_Q$ explicit and allows us to build up 
an expansion in $\Lambda_{\rm QCD}/m_Q$ by matching the
transition operator ${\cal T}$ in QCD
onto an expansion in terms of Heavy Quark Effective Theory 
(HQET) operators~\cite{Mannel:1991mc,Manohar:1997qy}.

Generally, the HQE for semileptonic weak decays
is written as~(e.g.~\cite{Benson:2003kp})
\begin{eqnarray} \label{rate-0}
\Gamma (B \to X_c \ell \bar{\nu}_\ell) =
\Gamma^0 |V_{cb}|^2 \left[ a_0(1 + \frac{\mu_\pi^2}{2m_b^2})
 +  a_2 \frac{\mu_G^2}{2m_b^2}  
 + \frac{a_D \rho_D+a_{LS} \rho_{LS}}{2m_b^3}  
+\ldots\right]  \nonumber
\end{eqnarray} 
where 
$\Gamma^0={\rm G_F^2} m_b^5/(192 \pi^3)$
and $m_b$ is the $b$-quark mass. 
The coefficients $a_i$, $i=0,2,D,LS$ depend on the ratio $m_c^2/m_b^2$, while 
$\mu_\pi^2$, $\mu_G^2$, $\rho_D$ and $\rho_{LS}$ are forward matrix elements 
of local operators, usually called HQE parameters. 

The precise definition of the appropriate mass parameter for the heavy quark 
field is of utmost importance for the precision of the predictions of the HQE 
and is thus extensively discussed in the literature, see e.g.~\cite{Penin:1998wj}. 
The HQE parameter $\mu_\pi^2$ is the kinetic energy parameter for the $B$-meson in
HQE, $\mu_G^2$ is the chromo-magnetic parameter. The term $\rho_{LS}$ contains
the spin-orbital interaction and $\rho_D$ is the Darwin term which is of our main
interest in the present paper.

The power suppressed terms are becoming important
phenomenologically as the precision of experimental data continues to 
improve. The coefficients $a_i$ have a perturbative expansion in the 
strong coupling constant $\alpha_s (m_Q)$. The leading coefficient $a_0$ 
is known analytically to~${\cal O}\left(\alpha_s^2\right)$ 
precision in the massless limit for the final 
state quark~\cite{vanRitbergen:1999gs}.
At this order the mass corrections have been  analytically accounted 
for the total width as an expansion in the mass of the final
fermion~\cite{Pak:2008qt} and for the differential distribution
in~\cite{Melnikov:2008qs}. 

The coefficient of the kinetic energy parameter is linked to the coefficient 
$a_0$ by reparametrization invariance (e.g.~\cite{Becher:2007tk}). 
The NLO correction to the coefficient of the chromo-magnetic parameter $a_2$
has been investigated in~\cite{Alberti:2013kxa} 
where the hadronic tensor has been computed analytically and the total decay rate has been then obtained by direct numerical
integration over the phase space.
This calculation allows for the
application of different energy/momentum cuts in the phase space
necessary for the accurate comparison with experimental data. 

The NLO strong interaction
$\alpha_s$ correction to the chromo-magnetic coefficient
$a_2$ in the total decay rate
has been analytically computed
in~\cite{Mannel:2014xza,Mannel:2015jka}.
The techniques of ref.~\cite{Mannel:2014xza,Mannel:2015jka} allow
also for an analytical computation of various moments in the hadronic
invariant mass or/and that of the lepton pair.
In the present paper we give the NLO result for the coefficient $a_D$ 
of $\rho_D$ in the analytical form retaining the full
dependence on the charm quark mass. 

\section{NLO for Darwin term \boldmath $\rho_D$ \unboldmath:
  Calculation and Results}
\label{sec:HQE}
In this section we describe the actual computation of the coefficient $a_D$
of the Darwin term. The present calculation follows the techniques used earlier
for the determination of NLO corrections to
the chromo-magnetic operator coefficient in the total
width~\cite{Mannel:2014xza,Mannel:2015wsa,Mannel:2015jka}.
Here we give a brief outline of the calculational setup,
for details of the techniques, see~\cite{Mannel:2015jka}.

We consider a normalized transition operator $\tilde {\cal T}$ defined by 
\begin{equation}
{\rm Im} {\cal T}=
\Gamma^0|V_{cb}|^2\tilde {\cal T}
\, .
\end{equation}
The heavy quark expansion for the rate is constructed by using
a direct matching from QCD to HQET 
\begin{equation}\label{eq:HQE-1}
\tilde {\cal T}=
C_0 {\cal O}_0  + C_v\frac{{\cal O}_v}{m_b}
+ C_\pi \frac{{\cal O}_\pi}{2m_b^2}  
+ C_G\frac{{\cal O}_G}{2m_b^2}
+ C_D\frac{{\cal O}_D}{2m_b^3}
\end{equation}
where we retain only the Darwin term in the $1/m_Q^3$ order.
The local operators~${\cal O}_i$ in the expansion~(\ref{eq:HQE-1}) are
ordered by their dimensionality
\begin{eqnarray} 
&& {\cal O}_0 =\bar{h}_v h_v  \qquad\qquad \mbox{(dimension three in mass units),} \\
&& {\cal O}_v =\bar{h}_v v\pi h_v \qquad\qquad \mbox{(dimension four in mass units),} \\ && {\cal O}_\pi =\bar{h}_v\pi_\perp^2 h_v 
\qquad\qquad \mbox{(dimension five in mass units),} \\ 
&& {\cal O}_G ={\bar h}_v\sigma^{\mu\nu}G_{^{\mu\nu}}h_v
   \qquad\qquad \mbox{(dimension five in mass units),}  \\
&& {\cal O}_D={\bar h}_v[\pi_\perp^\mu,
[\pi_\perp^\mu,\pi v]] h_v \qquad\qquad \mbox{(dimension six in mass units).} 
\end{eqnarray} 
Here the field $h_v$ is the heavy quark field the dynamics of which is given by the 
QCD Lagrangian expanded to order $1/m_Q^3$.  Furthermore, 
$\pi_\mu = i D_\mu$ is the covariant derivative of QCD and 
$\pi^\mu =v^\mu (v\pi)+\pi^\mu_\perp$. 
The coefficients  $C_0$, $C_v$, $C_G$, $C_D$ of the operators
are obtained by matching the appropriate matrix elements between 
QCD and HQET.

After taking the forward matrix element with the $B$-meson state
one can use the HQET equations of motion for the field $h_v$
in order to eliminate the operator ${\cal O}_v$.
We note that, in general, there is an additional operator 
${\cal O}_5=\bar{h}_v (v\pi)^2 h_v $ in the complete basis at dimension 
five, however it will be of higher order in the HQE
after using equations of motion of HQET.

Note that one can use the full QCD fields for the heavy quark
expansion afterwards~\cite{Mannel:2018mqv}.
It is convenient to choose the local operator  
$\bar{b}\slashed{v} b$ defined in full QCD as a leading term of 
the heavy quark expansion~\cite{Manohar:1993qn}
as it is absolutely normalised and provides a
direct correspondence to the quark parton model as the leading order of the HQE. 

We note that (\ref{eq:HQE-1}) is an operator relation and hence 
the coefficient functions $C_i$ are independent of any external states.  
Thus these states can be freely chosen as long as they comply with 
the requirements of HQE. Thus, for the matching of QCD to HQET we 
can chose external states built from gluons and heavy quarks, and 
the matching procedure consists in computing matrix elements of the relation~(\ref{eq:HQE-1}) with partonic states built from quarks and gluons. 

\begin{figure}[h!]
\centering 
\includegraphics[width=0.35\textwidth]{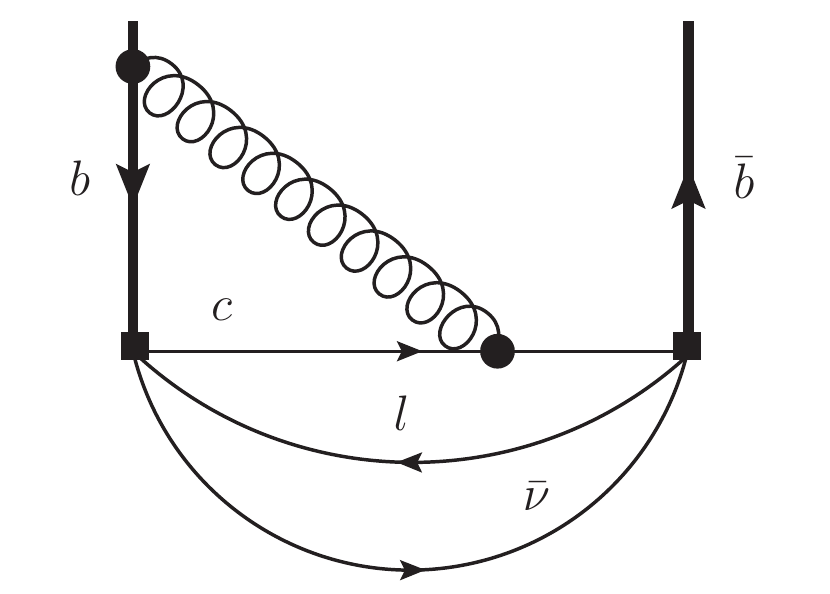}
\qquad
\includegraphics[width=0.35\textwidth]{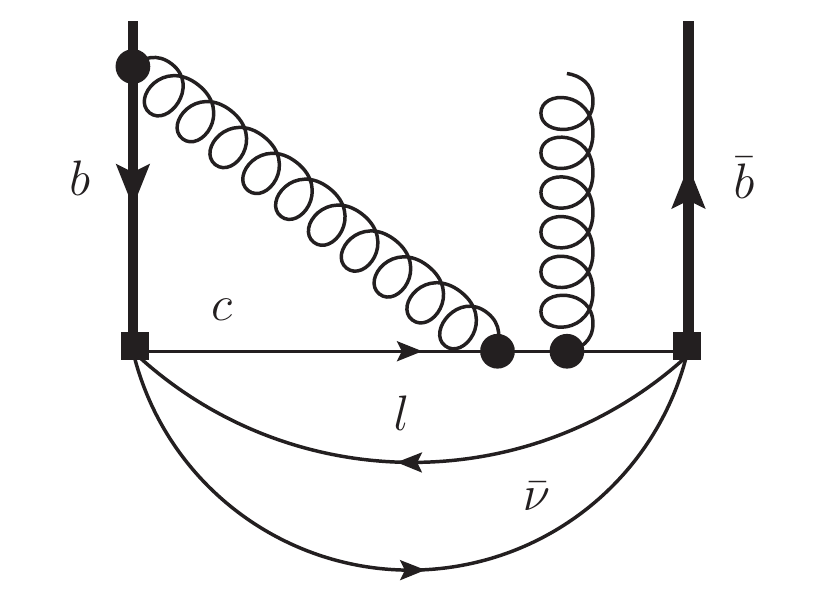}
\caption{\label{fig:diags-1}
Diagrams for the 
contribution at NLO level, 
(left) - partonic type, right - power correction 
type with an insertion of an external gluon}
\end{figure}

The coefficient function $C_0$ determines the total width of the heavy quark 
and, at the same time, the leading power contribution to the total width of 
a bottom hadron within the HQE.  At NLO the contributions to the 
transition operator $\tilde{\cal T}$ in~(\ref{eq:trans_operator})
are represented by three-loop Feynman diagrams shown in (\ref{fig:diags-1}).

The leading order 
result is given by two-loop Feynman integrals of a
simple topology
-- the so called sunset-type diagrams~\cite{Groote:1998wy,we-annals}, while 
at the NLO level 
one has to evaluate three-loop integrals with massive 
lines due to the massive $c$- and $b$-quarks.
In Fig.~\ref{fig:diags-1} we show a typical three-loop diagram
for the power corrections in the heavy quark expansion. 

We use dimensional regularization for
both ultraviolet and infrared singularities. We used the systems of symbolic
manipulations REDUCE~\cite{reduce} and Mathematica~\cite{Mathe}
with special codes written for the calculation. 
For reduction of integrals to master integrals
the program LiteRed~\cite{Lee:2013mka} 
was used.
The master integrals have been then computed directly.
Some of them have checked
with the program HypExp~\cite{Huber:2007dx}.

For the Darwin term one takes an
amplitude of quark to quark-gluon scattering and
projects it to an HQET operator.
We choose a momentum $k$ gluon and
take the structure $(\epsilon v) k_\perp^2$.
There are several operators in HQET that can have such a
structure, for instance, ${\bar h}_v(\pi v)\pi_\perp^2h_v$. This
operator is irrelevant because it is of higher
power on shell.
One disentangles the mixing of such operators with the Darwin term by
using two quark momenta $k_1$ and $k_2$ and pick up the structure
$(k_1k_2)$ that emerges in the coefficient of the Darwin term. 
The other operators can have $k_1^2$ or  $k_2^2$ structure.
The coefficient $a_D$ is defined in
front of the meson matrix element. After taking the matrix element one
can use equation of motion of HQET to reduce the number of the
operators in the basis. One more conventional step is to trade the
leading order operator ${\bar h}_vh_v$ for the QCD operator
${\bar b}\slashed{v}b$ that provides correspondence to the parton model.

The final expression for the coefficient $a_D$ is then
\begin{eqnarray}\label{eq:rhoD-LO-final}
  a_D&=&2 (C_D + \frac{3}{4} (C_v C_D^{HQET} - C_0 C_D^{bvb}))
\end{eqnarray}
where $C_D^{HQET}$ is the NLO coefficient of the operator $O_D$ in HQET
Lagrangian,
and $C_D^{bvb}$ is the NLO coefficient of the operator $O_D$
in the expansion of ${\bar b}\slashed{v}b$.
At the LO we find
\begin{eqnarray}\label{eq:rhoD-LO-direct}
  a_D^{LO}=-5 r^4-8 r^3+24 r^2+36 r^2 \log(r)-88 r+48 \log(r)+77
\end{eqnarray}
where $r=m_c^2/m_b^2$ that agrees with~\cite{Gremm:1996df}.
The coefficient contains 
a logarithmic singularity $\log (r)$ at small $r$.
This singularity reflects the mixing to hidden/intrinsic
charm contribution~\cite{Bigi:2009ym}.
At higher powers even more singular terms (like $1/r$) can
appear~\cite{Mannel:2010wj}.
The matching is performed
by integrating out the charm quark simultaneously
with the hard modes of the $b$-quark.
This means that we treat $m_c^2/m_b^2$ as a number   
fixed in the limit $m_b \to \infty$, 
and therefore our results cannot be used to extrapolate to the limit
$m_c\to 0$. 

An important check of a loop computation 
consists in verifying the cancellation of poles after performing the appropriate 
renormalization of the physical quantity in question. Since in the case at hand this
is quite delicate, we briefly discuss the renormalization of the $\rho_D$ coefficient
at NLO within our computation.

We single out the pole contribution to the NLO coefficient in the form
\begin{eqnarray}\label{eq:rhoD-NLO-poles}
C^{NLO}_D=\frac{\alpha_s(m_b)}{4\pi}
\left(\frac{1}{\epsilon} C^{NLO-pol}_D
+ C^{NLO-fin}_D\right)\, .
\end{eqnarray}
The contribution to the coefficient
$C^{NLO-pol}_D$ 
from one-particle irreducible diagrams
reads
\begin{eqnarray}\label{eq:rhoD-NLO-poles}
C^{NLO-pol}_D &=& C_A
                             \left(-\frac{17 r^4}{3}+\frac{16 r^3}{3}
                      - 28 r^2+36 r^2 \log (r)+\frac{32 r}{3}
                      + 16 \log(r)+\frac{53}{3}\right)  \nonumber  \\
  &+&C_F \left(-\frac{1181
   r^4}{8}+207 r^3+87 r^2+\frac{285}{2} r^2 \log
   (r)-419 r+72 \log (r)+\frac{2181}{8}\right)\, .        \nonumber
\end{eqnarray}
The pole part of the coefficient $C^{NLO-pol}_D $
contains different functional dependencies on $r$, and it is 
instructive to see how the cancellation works in the present  case.

The proper cancellation of these poles requires to consider the mixing between
HQE operators of different dimensionality which is known to be possible in
HQET~\cite{Falk:1990pz,Bauer:1997gs,Finkemeier:1996uu,Balzereit:1996yy,Blok:1996iz,Lee:1991hp}.
The anomalous dimensions of the operators
are numbers independent of $r$ while the functions of $r$ appearing in  
$C_D$ should cancel in the renormalization of the coefficient.  
This is a rather strong restriction because only the coefficient functions 
of the lower power operators (which are basically $C_0(r)$ and $C_v(r)$) 
can be used for the pole cancellation in $C_D$. 
The leading order
$C_G$ coefficient is proportional to
$C_0^{LO}(r)$.
Indeed, the explicit expression reads
\begin{eqnarray} \label{eq:cG-LO-dir}
C_G^{LO}=2 -16r -24 r^2\ln(r)+16 r^3 -2 r^4 = 2 C_0^{LO}
\, .
\end{eqnarray} 
These properties of HQE are important for the implementation
of the renormalization procedure of the  Darwin-term coefficient.

The operator ${\cal O}_\pi$ from HQE after the insertion of one more
${\cal O}_\pi$ from the Lagrangian can mix with ${\cal O}_D$ that produces the pole
structure proportional to $C_0(r)$
\begin{eqnarray}\label{eq:rhoD-NLO-poles-Ok-mix}
  {\cal O}_\pi^R={\cal O}_\pi^B
  +\gamma_{\pi D}\frac{\alpha_s}{4\pi\epsilon}\frac{1}{m_b}{\cal O}_D\, .
\end{eqnarray}
The relation~(\ref{eq:rhoD-NLO-poles-Ok-mix})
means that the $ghh$ vertex computed in perturbation theory within HQET with one
insertion of ${\cal O}_\pi$ gets a contribution from higher powers of
the HQET Lagrangian (see, e.g.~\cite{Falk:1990pz}).
By the same token the operator ${\cal O}_G$ from HQE after the insertion of 
one more ${\cal O}_G$ from the Lagrangian can mix with ${\cal O}_D$
\begin{eqnarray}\label{eq:rhoD-NLO-poles-Om-mix}
  {\cal O}_G^R={\cal O}_G^B
  +\gamma_{GD}\frac{\alpha_s}{4\pi\epsilon}\frac{1}{m_b}{\cal O}_D
\end{eqnarray}
that produces the pole structure proportional to $C_0(r)$ again because 
of eq.~(\ref{eq:cG-LO-dir}).
The cross-insertions (${\cal O}_G$ from HQE
to ${\cal O}_\pi$ from the Lagrangian and vice
versa) renormalize the spin-orbit operator  
at the order $1/m_b^2$. This type of mixing is known for a long time
from computation of  $1/m_b^2$ running of coefficients of HQET
Lagrangian.

In the literature 
the renormalization is considered often for the static heavy fields
when the contributions of reiterated terms in the HQET Lagrangian are
accounted for through the bi-local
operators~\cite{Balzereit:1996yy,Blok:1996iz}. We consider the standard
approach and treat higher order terms as perturbations
(see,~\cite{Finkemeier:1996uu,Bauer:1996ma}). 
In our case the inclusion of these mixings
does not suffice to cancel all the poles in $C_D$ 
as there are other structures than $C_0(r)$ necessary.  
The operator ${\cal O}_v$ can mix with double insertions of higher
dimensional terms, i.e.
\begin{eqnarray}\label{eq:rhoD-NLO-poles-Om-mix}
  {\cal O}_v^R={\cal O}_v^B
  +\gamma_{vD}\frac{\alpha_s}{4\pi\epsilon}\frac{1}{m_b^2}{\cal O}_D
\end{eqnarray}
and the counterterm proportional to ${\cal O}_D$ emerges from two
insertions of the operators $ {\cal O}_\pi$.
The mixing matrix $\gamma_{vD}$
is unknown. But the effect of such a mixing
leads to the appearance of the coefficient $C_v(r)$ in
the expression for the poles.
One can now fit the pole function with two entries $C_0(r)$
and $C_v(r)$.  

Thus we infer the corresponding mixing anomalous dimensions
and find that the combination
\begin{eqnarray}\label{eq:rhoD-NLO-poles-guess}
(-C_A + \frac{23}{8}C_F)C_v(r) - (\frac{5}{4}C_A
  + \frac{31}{8}C_F)C_0(r)
\end{eqnarray}
cancels the poles in both color
structures $C_F$ and $C_A$ for the entire $m_c$ dependence.
The solution in eq.~(\ref{eq:rhoD-NLO-poles-guess}) is unique.
The presence of the coefficient $C_v$ means an admixture to the
operator ${\cal O}_v$.
At this level it is impossible to 
confirm the two mixings as the mixing matrices are still
not uniquely given in the
literature and $\gamma_{vD}$ is completely new.
An independent computation of mixing matrices could be a useful
check of our
computation.
Because the term $r^2$ is present only in the mixing with ${\cal O}_v$
one can extract  $\gamma_{vD}$. But it is impossible to separate
$\gamma_{mD}$ and $\gamma_{kD}$ as only their sum is
extracted with our current 
method.  

Thus we arrive at a finite coefficient for $\rho_D$, 
the analytical expression for NLO correction to $a_D$ is given in the
Appendix. Here we discuss the numerical impact of our result. With 
$\alpha_s$ normalized at $m_b$ and for
$r=m_c^2/m_b^2=0.07$
one finds
\begin{eqnarray}
  a_{D}&=&-57.159+\frac{\alpha_s(m_b)}{4\pi}(-56.594 C_A + 408.746 C_F)
           \nonumber\\
      &=&-57.159+\frac{\alpha_s(m_b)}{4\pi}(375.213)  \nonumber\\
  &=&-57.159(1-\frac{\alpha_s}{4\pi}6.564\ldots)\label{Final0}
\end{eqnarray}
For $\alpha_s(m_b)=0.2$
\begin{eqnarray}
  a_{D}&=&-57.159(1-0.10)
           \label{Final}
\end{eqnarray} 
the NLO contribution shifts the $\rho_D$ coefficient by 10\%. 

\section{Discussion}
The technical details of the calculation will be discussed in a 
more detailed paper, where we also plan to calculate moments of various 
distributions. However, the result presented here already have a few 
interesting consequences. 

\begin{figure}[!htb]
    \includegraphics[scale=0.65] {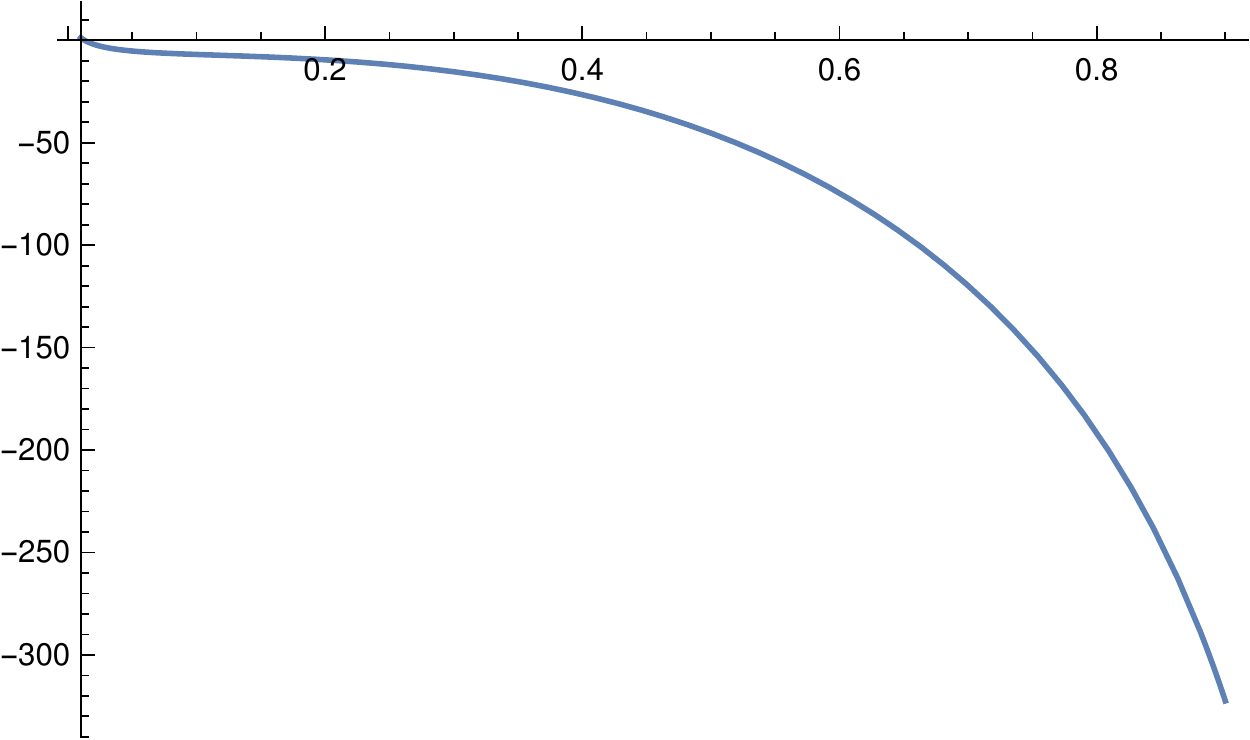}
    \hfill \includegraphics[scale=0.65] {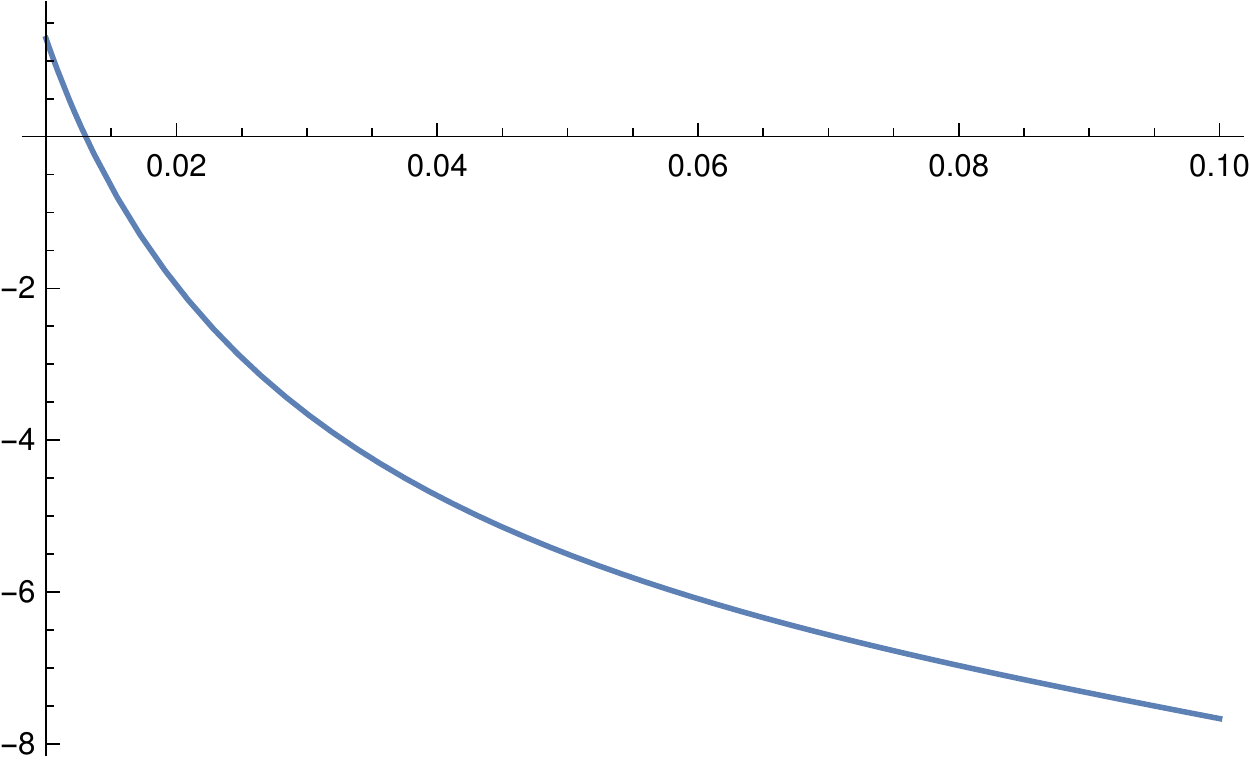}     
    \caption{\label{FullRange} Mass dependence of the NLO Coefficient
      of $\rho_D$.
      Left panel: The ratio~(\ref{Ratio})
    over the full range of $r$, right panel: The ratio of the NLO
    coefficient
    to the LO one in the physically 
    interesting region.  }
     \end{figure}

The first remark concerns the dependence on the mass of the charm quark
which appears in the ratio $r = m_c^2/m_b^2$. This ratio is kept at a fixed 
value as $m_b, m_c \to \infty$ and the behavior of the coefficients 
close to $r=0$ is given by 
\begin{eqnarray} 
a_D^{LO} &=& - 20 (1-r)^4 + \ldots  \\
a_D^{NLO} &=& - C_F 8 (1-r)^3 (29 + 312 \ln(1-r)) + \cdots
\end{eqnarray}
Note that the behaviour of the coefficient
at the border of the available phase space depends on the
definition of the $c$-quark mass. Here we use $\MSbar$ mass.
Ii Fig.~\ref{FullRange} (left panel) we plot the dependence on $r$ in the
full kinematically allowed region 
range $0 \le r \le 1$. We show the ratio 
\begin{equation} \label{Ratio}
\frac{a_D^{NLO} (1-r)}{a_D^{LO}}
\end{equation} 
while the right panel of Fig.~\ref{FullRange} focuses on the physical region 
around $r = 0.07$. 

The plots show that the mass dependence in the physical region is weak, while 
it is sizable over the full range. As we discussed above, the massless
limit cannot 
be taken, since in the case of a $b \to u$ transition additional operators have to be 
taken into account. Nevertheless at small $r$ the NLO corrections become even smaller 
and have a zero at $r \sim 0.005$. A similarly strong dependence has been observed 
also for the QCD corrections in the coefficient
of $\mu_G^2$ \cite{Mannel:2015jka}.   

Although the corrections are not untypically large, they will have 
a visible impact on the determination of $V_{cb}$. This is 
mainly due to the fact that the coefficient in front of $\rho_D$ in 
the total rate is quite large, see (\ref{Final}). While a detailed 
analysis will require to repeat the combined fit as e.g.
in~\cite{Gambino:2013rza} 
we may obtain a tendency from an approximate 
formula given in eq.~(12) in this paper. According to (\ref{Final}) the NLO correction 
corresponds to a reduction of the contribution of $\rho_D$ by 10\%, thus eq.~(12)
of~\cite{Gambino:2013rza} implies a shift in the central value of 
\begin{equation}
\frac{\Delta V_{cb}}{V_{cb}} = - 0.3\%  \, ,  
\end{equation} 
which is about a third of the current theoretical uncertainty. 

However, parametrically this correction is of the same size as the yet unknown
corrections of order $\alpha_s^2 \Lambda_{\rm QCD}^2 / m_b^2$ and $\alpha_s^3$, 
which would need to be included in a full analysis up to order 
$\alpha_s \Lambda_{\rm QCD}^3 / m_b^3$. We note in passing that the corrections
$\alpha_s \rho_{LS}^2$ are not needed, since these are included in the known 
$\alpha_s \mu_G^2$ contributions~\cite{Fael:2018vsp}.
Nevertheless, the contribution
of $\rho_D$ is significant due to the large coefficient in front of $\rho_D$ 
and hence we expect that the impact of this correction is largest.  

\vskip 0.5cm
{\bf Acknowledgments}\\
\noindent
This research was supported by the Deutsche Forschungsgemeinschaft 
(DFG, German Research Foundation) under grant  396021762 - TRR 257 
``Particle Physics Phenomenology after the Higgs Discovery'' 

\section{Appendix}
The matching coefficient of the ``operator'' $\rho_D$ in the HQET
Lagrangian in NLO at $\mu=m_b$
gets a correction~\cite{Manohar:1997qy}
\begin{eqnarray}
  1+\frac{\alpha_s(m_b)}{4\pi}2C_A\, .
\end{eqnarray}
After using equation of motions the final 
$\rho_D$-coefficient is expressed through the coefficient of
the relevant operator in HQE (direct contribution) and the
contributions due to HQET Lagrangian
and the choice of the full QCD operator at the leading
power
through the relation
\begin{eqnarray}
  a_{D}=2\left\{C_D^{dir}
  +\frac{3}{4} C_v \left(1+\frac{\alpha_s}{4\pi}2C_A\right)
  -\frac{3}{4}C_0\right\}\, .
\end{eqnarray}
At LO one obtains
\begin{eqnarray}
a_{D}^{LO}=-5 r^4-8 r^3+24 r^2+36 r^2 \log (r)-88 r+48 \log (r)+77
\end{eqnarray}
that agrees with~\cite{Gremm:1996df}.

We write the coefficient of $\rho_D$ term after taking matrix elements
as
\begin{eqnarray}
a_D=  a_{D}^{LO}+\frac{\alpha_s}{4\pi}a_{D}^{NLO}
\end{eqnarray}
and then
\begin{eqnarray}
a_{D}^{NLO}=a_{D}^{NLO,cf}C_F+a_{D}^{NLO,ca}C_A  \, .
\end{eqnarray}

The $C_F$ color part reads at NLO
  \begin{eqnarray}
&&a_{D}^{NLO,cf}=
\left(-1776 r^{5/2}-\frac{6464 r^{3/2}}{3}-144 r^4+120
   r^2-\frac{880 \sqrt{r}}{3}\right)
   \text{Li}_2\left(-\sqrt{r}\right)+\nn  \\
    &&\left(3408
   r^{5/2}+5312 r^{3/2}-144 r^4+120 r^2+880
   \sqrt{r}\right)
       \text{Li}_2\left(\sqrt{r}\right)\nn  \\
    &&+\left(168 r^4+256
   r^3-1304 r^2+928 r+424\right)
       \text{Li}_2\left(\frac{r-1}{r}\right)\nn \\
    &&+\left(-408
   r^{5/2}-\frac{2368 r^{3/2}}{3}-192 r^4-32 r^3+144
       r^2-\frac{440 \sqrt{r}}{3}+80\right) \text{Li}_2(r)\nn   \\
    &&-648
   \pi ^2 r^{5/2}-\frac{2800}{3} \pi ^2 r^{3/2}+44 \pi ^2
   r^4-\frac{1615 r^4}{48}+\frac{16 \pi ^2
       r^3}{3}-\frac{123184 r^3}{45}\nn   \\
    &&-34 \pi ^2
       r^2-\frac{363827 r^2}{180}\nn   \\
    &&+\left(164 r^4+\frac{880
   r^3}{3}-\frac{443 r^2}{3}+584 r+140\right) \log
       ^2(r)\nn  \\
    &&+\left(\frac{18677 r^4}{15}-\frac{54296
   r^3}{45}+\frac{1648 r^2}{9}+\frac{392}{15
   r^2}+\frac{6424 r}{3}-\frac{4496}{45
       r}-\frac{20603}{9}\right) \log (1-r)\nn \\
    &&+\left(\left(1296
   r^{5/2}+\frac{5600 r^{3/2}}{3}+\frac{880
   \sqrt{r}}{3}\right) \log
       \left(1-\sqrt{r}\right)\right.\nn   \\
    &&+\left(-1296 r^{5/2}-\frac{5600
   r^{3/2}}{3}-\frac{880 \sqrt{r}}{3}\right) \log
       \left(\sqrt{r}+1\right)\nn \\
    &&-\frac{2093 r^4}{60}+\frac{34466
       r^3}{45}+\frac{41815 r^2}{18}\nn  \\
    &&+\left(-296 r^4-\frac{352
   r^3}{3}-\frac{6904 r^2}{3}-576 r-\frac{1912}{3}\right)
       \log (1-r)\nn    \\
    &&\left.+\frac{3752 r}{3}-\frac{2096}{3}\right) \log
   (r)+\frac{220204 r}{45}-\frac{440 \pi ^2
       \sqrt{r}}{3}\nonumber \\
    &&+\frac{392}{15 r}-\frac{40 \pi
   ^2}{3}-\frac{91603}{720}
\end{eqnarray}
Here $\text{Li}_2(z)$ is a dilogarithm.

The new master integral $N_m$ has appeared compared to
our previous results. The relevant combination turns out to be
$(N_p+N_m)/2$, and both $N_p$ and $N_m$ should be evaluated
at LO in $\epsilon$-expansion.  

A similar expression for the $C_A$ color part of the NLO
coefficient reads
\begin{eqnarray}
&&a_{D}^{NLO,ca}=\left(-152 r^{5/2}-\frac{400 r^{3/2}}{3}-336 r^2+\frac{280
   \sqrt{r}}{3}\right)
   \text{Li}_2\left(-\sqrt{r}\right)\nn \\
  &&+\left(456 r^{5/2}+400
   r^{3/2}-336 r^2-280 \sqrt{r}\right)
     \text{Li}_2\left(\sqrt{r}\right)+\nn \\
  &&\left(-208
   r^3+\frac{712 r^2}{3}-360 r+216\right)
     \text{Li}_2\left(\frac{r-1}{r}\right)\nn \\
  &&+\left(-76
   r^{5/2}-\frac{200 r^{3/2}}{3}-24 r^3+72 r^2-72
     r+\frac{140 \sqrt{r}}{3}+24\right) \text{Li}_2(r)-\nn \\
  &&76
   \pi ^2 r^{5/2}-\frac{200}{3} \pi ^2 r^{3/2}+\frac{329
   r^4}{36}+4 \pi ^2 r^3+\frac{749 r^3}{9}+16 \pi ^2
     r^2+\frac{12941 r^2}{45}+\nonumber \\
  &&\left(-\frac{172
   r^3}{3}+\frac{268 r^2}{3}-216 r+40\right) \log
     ^2(r)+\nonumber \\
  &&\left(48 r^4-\frac{10114 r^3}{45}+134
   r^2+\frac{238}{15 r^2}+\frac{280
   r}{3}-\frac{62}{r}-\frac{40}{9}\right) \log
     (1-r)+\nn \\
  &&\left(\left(152 r^{5/2}+\frac{400
   r^{3/2}}{3}-\frac{280 \sqrt{r}}{3}\right) \log
   \left(1-\sqrt{r}\right)+\left(-152 r^{5/2}-\frac{400
   r^{3/2}}{3}+\frac{280 \sqrt{r}}{3}\right) \log
     \left(\sqrt{r}+1\right)\right.\nonumber \\
  &&\left.-26 r^4+\frac{11794 r^3}{45}-673
   r^2+\left(\frac{136 r^3}{3}-144 r^2+128
   r-\frac{320}{3}\right) \log (1-r)+\frac{508
     r}{3}-\frac{1060}{3}\right) \log (r)+\nonumber \\
  &&12 \pi ^2
   r-\frac{2009 r}{15}+\frac{140 \pi ^2
   \sqrt{r}}{3}+\frac{238}{15 r}-4 \pi
   ^2-\frac{47137}{180}
\end{eqnarray}
Note that there is no $1/r$ singularity at small $r$.
The small $r$ expansion for the $C_F$ structure is 
\begin{eqnarray}
&&a_{D}^{NLO,cf}=\left(-72 \log ^2(r)-\frac{2096 \log (r)}{3}-84 \pi
   ^2-\frac{5815}{144}\right)-\frac{440 \pi ^2
   \sqrt{r}}{3}\nn \\
  &&+\frac{1}{9} r \left(1080 \log ^2(r)+7896
   \log (r)-1392 \pi ^2+80111\right)+O\left(r^{3/2}\right)
\end{eqnarray}
and for the $C_A$ part is
\begin{eqnarray}
&&a_{D}^{NLO,ca}=
\left(-68 \log ^2(r)-\frac{1060 \log (r)}{3}-40 \pi
   ^2-\frac{7481}{36}\right)+\frac{140 \pi ^2
   \sqrt{r}}{3}\nn \\
    &&+r \left(-36 \log ^2(r)+\frac{740 \log
   (r)}{3}+72 \pi
   ^2-\frac{2134}{9}\right)+O\left(r^{3/2}\right)
  \end{eqnarray}

\end{document}